# Single crystal synthesis and magnetic properties of a Shastry-Sutherland lattice compound BaNd$_2$ZnS$_5$

Brianna R. Billingsley[1], Madalynn Marshall[2], Zhixue Shu[1], Huibo Cao[2], Tai Kong[1,3]

1. Department of Physics, University of Arizona, Tucson, Arizona, 85721
2. Neutron Scattering Division, Oak Ridge National Laboratory, Oak Ridge, Tennessee 37831, USA
3. Department of Chemistry and Biochemistry, University of Arizona, Tucson, Arizona, 85721

**Abstract**

Single crystals of a Shastry-Sutherland magnetic semiconductor, BaNd$_2$ZnS$_5$, were synthesized through a high-temperature solution growth technique. Physical properties were characterized by powder and single crystal x-ray diffraction, temperature- and field-dependent magnetization, and temperature-dependent specific heat measurements. BaNd$_2$ZnS$_5$ orders antiferromagnetically at 2.9 K, with magnetic moments primarily aligned within the *ab*-plane. Magnetic isothermal measurements show metamagnetic transitions at ~ 15 kOe for the [110] direction and ~ 21 kOe for the [100] direction. Estimated magnetic entropy suggests a double ground state for each Nd ion.

**Introduction**

Geometrical frustration is one of the most effective ways to induce magnetic frustration in solids, which is commonly achieved by incorporating a triangular motif in the magnetic ion sublattice [1]. A Shastry-Sutherland lattice (SS) represents one of the geometrical frustrated lattice types. The original SS model considers competing magnetic interactions between a nearest-neighbor (NN) and an alternating next-nearest-neighbor (NNN) in a two-dimensional square lattice [2]. Depending on the relative ratio of the two magnetic interactions and spin number, different magnetic ground states were predicted, including long-range ordering and quantum spin liquid [2]. In real materials, to fulfill the required alternation of NNN interaction in the SS model, every other square unit needs to be distorted diagonally, essentially forming a pattern that is constructed by squares and triangles. Fig. 1 illustrates an SS lattice, where *J* represents the NN interaction, also called interdimer interaction, and *J'* represents the NNN interaction, also called intradimer interaction. Materials will exhibit long-range magnetic ordering at a large *J/J'* and a dimer singlet state at a small *J/J'* [3,4]. The possibility of intermediate ground states, however, is under intensive theoretical investigation [5–14], and lacks sufficient experimental realization. The first example of an SS material is $SrCu_2(BO_3)_2$, where a gapped dimer singlet ground state was realized with a *J/J'* at ~ 0.6 [3,15,16]. More materials were later demonstrated to host an SS lattice, examples being metallic $Gd_2Ge_2Mg$ [17], $REB_4$ [18–20], and $RE_2Pt_2Pb$ [21] (*RE* = rare earth). Whereas long-range RKKY type interaction must be considered in a metallic compound, magnetism in a semiconducting or insulating SS lattice may involve simpler exchange interactions, thus maps closer onto the original SS model. As part of experimental exploration of different ground states of SS lattices, this study focuses on the single crystal study of a rare-earth-based semiconducting compound, $BaNd_2ZnS_5$.

The general chemical formula $BaRE_2TMCh_5$ (*TM* = transition metal, *Ch* = O, S) hosts many different materials [22–26]. For oxides, nearly the entire *RE* series is reported to exist when *TM* = Pd [22,27]. When *TM* = Zn or other 3d transition metals, the *RE* series will still form. However, the crystal structure changes from an SS lattice with a space group of *I*4/mcm for lighter *RE*s, to an orthorhombic lattice with a space group of *P*bnm for heavier *RE* elements, showing no SS lattice [28–30]. Such structural change sets up a limit on possible chemical tuning studies. The SS lattice can be restored for some heavier *RE* elements through high-pressure synthesis [31]. For sulfide, only the La-Nd versions are reported with an *a*-axis constant of ~ 7.9 Å and a *c*-axis constant of ~ 13.6 Å. Both sulfide lattice constants are larger than the oxide version, which is consistent with ionic size differences between oxygen and sulfur [32]. $BaNd_2ZnS_5$ crystallizes in the *I*4/mcm space group (#140) with a $Cs_3CoCl_5$ structure type. A schematic crystal structure drawing of $BaNd_2ZnS_5$ is shown in Fig. 1. Neodymium atoms occupy a single atomic site in the crystal structure, which forms a planar SS lattice bridged by sulfur atoms [24]. Zinc also occupies only one atomic site, forming a square lattice in the *ab*-plane. Each SS layer is separated from each other with a layer of barium and transition metal, coordinated by sulfur. The magnetic SS layers are thus separated from each other by ~ 6.8 Å. Within each SS plane, the long- and short-edge lengths in each triangular unit are about 4.151 Å (inter-dimer) and 3.596 Å (intra-dimer), respectively. In addition to the difference in inter-neodymium distance, the Nd-S-Nd angle is different for inter- and intra-dimer. As shown in Fig. 1b, the inter-dimer is channeled through three Nd-S-Nd links, each having an angle of ~ 90°. The intra-dimer is channeled through four Nd-S-Nd links, each having an angle of ~ 72°. For $BaNd_2ZnO_5$, the inter-dimer angle remains close to 90° while the intra-dimer angle is about 78° [33].

Previously, Ba*RE*$_2$*TM*S$_5$ (*RE* = La-Nd, Sm; *TM* = Mn, Fe, Co, Zn) had only been synthesized in a powder form, where magnetic transition metal exhibit antiferromagnetic ordering at ~ 60 K, and magnetic rare earth elements order at a much lower temperature below 5 K [24–26,34]. To study the magnetic behavior of the SS lattice, it is best to be free from other magnetic contributions. In this paper, we focus on $BaNd_2ZnS_5$, where the transition metal site is occupied by non-magnetic zinc. Measurements on $BaNd_2ZnS_5$ powder sample suggest an insulating state with an activation energy of 1.46 eV, and an antiferromagnetic phase transition at 2.8 K [34]. Here we report the single crystal synthesis and detailed anisotropic magnetic properties of $BaNd_2ZnS_5$.

**Experimental methods**

$BaNd_2ZnS_5$ single crystals were synthesized using the high temperature solution method [35]. The starting elements comprised of barium pieces (Alfa Aesar, 99.2%), neodymium pieces (Alfa Aesar, 99.5%), zinc shot (Alfa Aesar, 99.999%) and sulfur pieces (Alfa Aesar, 99.999%) were packed into an alumina Canfield Crucible Set [36] with a molar ratio of Ba:Nd:Zn:S = 5:4:2:19, which was then sealed in a silica tube under vacuum. Air sensitive barium and neodymium were handled in an argon glovebox. The ampoule was first heated to 430 °C over 3 hours, dwelled for 6 hours, then to 850 °C over 12 hours, where it dwelled for 5 hours, then finally heated to 1060 °C over 6 hours, and dwelled for 10 hours. After dwelling at 1060 °C, the ampoule was slowly cooled to 750 °C for decanting. The slow heating process is necessary to protect the alumina crucible from being attacked by barium and neodymium. Faster heating can result in a broken alumina crucible and silica ampoule. At 750 °C, significant sulfur vapor pressure is still present as evidenced by a light brown colored vapor found inside of the silica tube. Dark-brown, cubic-like single crystals of $BaNd_2ZnS_5$ can be collected from the growth crucible. Typical crystals are shown in the Fig. 2 inset on millimeter grid paper.

Room temperature powder x-ray diffraction (PXRD) was measured using a Bruker D8 Discover diffractometer, with a Cu Kα radiation (λ = 1.5406 Å). A selection of single crystals was ground into powder and evenly spread on a vacuum greased glass slide for measurement. PXRD data were analyzed using GSAS and the Le Bail method [37,38]. Crystalline orientation was determined by collecting single crystalline diffraction peaks using the powder diffractometer [39]. Single crystal x-ray diffraction (SXRD) measurements were carried out on a Rigaku XtaLAB PRO diffractometer using a Mo $K_α$ radiation, for a crystal of $BaNd_2ZnS_5$ at room temperature. Data collection and integrated was done using the Rigaku Oxford Diffraction CrysAlis Pro software [40] and the structural refinement was performed using a SHELXTL package [41,42]. The crystal structure was drawn using VESTA [43].

Anisotropic magnetization and specific heat data were measured using a Quantum Design (QD) physical property measurement system (PPMS) Dynacool (1.8-300 K, 0-90 kOe). Magnetization was measured using the vibrating sample magnetometer (VSM) function. Single crystalline samples were manually mounted on a silica sample holder in preferred orientation with GE varnish. Zero-field specific heat data were measured down to 1.8 K using the two-tau relaxation method. The specific heat of $BaLa_2ZnS_5$ single crystal was also measured to serve as the non-magnetic lattice contribution for the estimation of magnetic entropy in $BaNd_2ZnS_5$.

**Results and discussion**

PXRD data shown in Fig. 2 are in good agreement with reported crystal structure of $BaNd_2ZnS_5$ [24]. Non-magnetic Ba-S binary impurity phases can sometimes be found on the surface of single crystals, which can be mechanically removed. Facets of as-grown single crystals can be indexed by either (00*l*) or (*hh*0). A selection of observed single crystal diffraction peaks is shown in the lower panels of Fig. 2. Because BaNd$_2$ZnS$_5$ has a tetragonal crystal structure, [100] was aligned by a 45° in-plane rotation away from the [110] direction. Room temperature SXRD result is consistent with previously reported powder refinement result [24]. Detailed crystal structure refinement results are listed in Table 1-3. Recent neutron scattering data on BaNd$_2$ZnO$_5$ [33] and BaNd$_2$ZnS$_5$ [44] suggest that Nd local magnetic moments prefer to align along [110] or the equivalent direction [1 -1 0], breaking the in-plane magnetic isotropy. In terms of crystal structure, anisotropic displacement parameters listed in Table 3 show a small displacement anisotropy of Nd ions at room temperature, which shows the symmetry similarity with its in-plane magnetic anisotropy at low temperatures. Such coincidence may worth further investigation.

Fig. 3 shows the anisotropic, temperature-dependent magnetization of BaNd$_2$ZnS$_5$ measured at 10 kOe. The polycrystalline averaged magnetization was calculated via $\chi_{poly}$ = 1/3 $\chi_{001}$ + 1/3 $\chi_{100}$ + 1/3 $\chi_{110}$. Here $\chi$ is defined as *M*/*H*. At high temperatures, the magnetization is anisotropic with $\chi_{ab}$ > $\chi_c$, which is likely due to the crystal electric field (CEF) effect that is commonly seen in rare earth based compounds [45,46]. In-plane magnetic anisotropy is weak at 10 kOe and thus only $\chi_{100}$ is displayed. A Curie-Weiss fit ($\chi$ = C/(*T*-$\theta$)) to the high-temperature polycrystalline averaged data gives an effective moment of 3.3 $\mu_B$/Nd, close to the theoretical value of 3.62 $\mu_B$/Nd. Anisotropic Curie-Weiss temperatures are $\theta_{poly}$ = -8 K, $\theta_{ab}$ = -1 K, $\theta_c$ = -21 K, indicating primarily antiferromagnetic interactions. At low temperatures, a clear drop in magnetization indicates an antiferromagnetic ordering. Taking the peak in d($\chi T$)/d*T* [47] as a criteria, the magnetic ordering temperature is determined as ~2.7 K.

Metamagnetic transitions were observed below the antiferromagnetic ordering temperature. Fig. 4 shows anisotropic magnetic isotherm measurements at 1.8 K. Along the *c*-axis ([001] direction), the magnetization increases linearly with applied magnetic field up to 90 kOe. For in-plane orientations, the metamagnetic transition appears at ~ 15 kOe for the [110] direction and at ~ 21 kOe for the [100] direction. The metamagnetic transition fields were determined by the peak fields in d*M*/d*H* curves. The ratio of the metamagnetic phase transition fields between the two in-plane orientations is close to $\sqrt{2}$, indicating that an equivalent 15 kOe magnetic field along [110] is needed to induce the metamagnetic phase. This is consistent with the recent neutron scattering result, indicating magnetic moments are along [110] and equivalent directions [44]. Kinks in magnetization appear at ~ 0.35 μ$_B$/Nd for [110] and at ~0.51 μ$_B$/Nd for [100]. Because 1.8 K is close to the magnetic ordering temperature, it is possible that these magnetization kinks may appear more plateau-like at lower temperatures, similar to other SS lattice compounds. Note that the d$M$/d$H$ curve for [100] (Fig. 4 inset) has a relatively rounded peak shape compared to that along [110], which may suggest the existence of a small magnetization plateau. The magnetization for both in-plane orientations appears to be saturating at high magnetic field. Extrapolating the linear magnetization curve above 60 kOe back to zero field gives a moment size of ~ 1.2 μ$_B$ for [110] and ~ 1.8 μ$_B$ for [100]. Such in-plane magnetic anisotropy likely result from a combined influence of CEF effect and applied magnetic field [48]. The anisotropic magnetization data on diluted, BaLa$_{1.86}$Nd$_{0.14}$ZnS$_5$ single crystals are shown by dashed lines in Fig. 4. The doping ratio was measured by scanning electron microscopy. The single ion results obtained on the diluted sample also show an in-plane magnetic anisotropy at high magnetic fields, which accounts for only part of the observed magnetic anisotropy in the pure BaNd$_2$ZnS$_5$. Additional in-plane anisotropy in pure BaNd$_2$ZnS$_5$ may come from a change in CEF scheme due to a strong mean field in the magnetically ordered state.

Fig. 5 shows the zero-field specific heat ($C_p$) data of BaNd$_2$ZnS$_5$. A λ-shaped peak suggests a second-order phase transition at 2.9 K, which is consistent with the antiferromagnetic ordering observed in the magnetization data and in a previous report [34]. Magnetic specific heat ($C_{mag}$) can be estimated by subtracting the $C_p$ of BaLa$_2$ZnS$_5$ as the lattice contribution, as shown by the red solid line in Fig. 5. The magnetic entropy of BaNd$_2$ZnS$_5$ reaches close to $R$ln2 per Nd ion above 10 K, which indicates a doublet CEF ground state of Nd in BaNd$_2$ZnS$_5$. At high temperature, a broad dome in $C_{mag}$ centered around 60 K is likely a Schottky anomaly due to CEF excitation. Because Nd$^{3+}$ is a Kramer's ion, $C_{mag}$ can be fitted with five doublets Schottky excitation as:

$$C_{mag} = \frac{2R}{T^2} \frac{\left(\sum_{i=2}^{i=5} \varepsilon_i^2 e^{-\varepsilon_i/T}\right) \left(1 + \sum_{i=2}^{i=5} e^{-\varepsilon_i/T}\right) - \left(\sum_{i=2}^{i=5} \varepsilon_i e^{-\varepsilon_i/T}\right)^2}{\left(1 + \sum_{i=2}^{i=5} e^{-\varepsilon_i/T}\right)^2}$$

Here $\varepsilon_i$ is the energy of each CEF doublet in the unit of kelvin; $R$ is the natural gas constant; $T$ is temperature; and the pre-factor 2 comes from two Nd ions per formula unit. The ground state $\varepsilon_1$ is assumed to be zero. The fitted curve is shown by the dotted magenta line in the right inset of Fig. 5. The broad dome around 60 K is mostly due to a first excited state at ~100 K and two close-by doublets at ~200 K. Another doublet may situate at higher energies. Better estimation on higher-lying CEF levels will require $C_{mag}$ data at higher temperatures. Comparing to the previous polycrystalline report where only one doublet was fitted at 197 K, the current estimation of two close-by doublets at ~200 K is closer to the scenario found in $BaNd_2ZnO_5$ where 3 doublets were reported to exist within 5 meV. More detailed CEF level determination via inelastic neutron scattering will be published elsewhere.

The fractional magnetization plateau in SS model is of great theoretical interest. Considering magnetic coupling beyond the original NN and NNN interaction can bring in additional fractions other than the commonly predicted 1/3 $M_{sat}$ for 2D [5]. Whereas most models consider Ising spin, the current planar magnetization would be more suitably described using an XY model. The planar magnetic anisotropy also posts difficulty in defining the fraction of magnetization plateau at the metamagnetic transition because the size of the local moment originates from the interplay between CEF level hybridization and applied magnetic field [48]. Using the magnetization values shown in Fig. 4, the metamagnetic phase at ~ 20 kOe roughly corresponds to 1/3 - ¼ of the saturation value for each in-plane orientation.

The physical properties of $BaNd_2ZnS_5$ are similar to that of recently reported $BaNd_2ZnO_5$ [33,49], which orders antiferromagnetically at a lower temperature of 1.65 K. Given that the Nd-Nd distance in $BaNd_2ZnS_5$ is longer by ~ 3.5% for intradimer and ~ 18% for interdimer compared to $BaNd_2ZnO_5$, thus reducing dipole-dipole interaction, sulfur bridging atoms may be facilitating a stronger Nd-Nd interaction than oxygen atoms. Continuing the current trend in crystal structure, replacing sulfur with larger chalcogen atoms such as selenium will further expand the lattice. Additionally, the ratio between intradimer and interdimer distance may continue to decrease, thus favoring the formation of a singlet dimer ground state if an antiferromagnetic exchange interaction is considered. The CEF ground state is also a doublet in $BaNd_2ZnO_5$ with an ordered moment of 1.9 $\mu_B$. This is comparable to the current observed saturation moment of ~1.8 $\mu_B$ for $BaNd_2ZnS_5$. Zero-field neutron scattering measurement on $BaNd_2ZnO_5$ and $BaNd_2ZnS_5$ revealed an antiferromagnetic structure with in-plane orthogonally arranged ferromagnetic dimers [33,44]. This is similar to the theoretical picture of a Neel state [7], except without catching the orthogonal arrangement of the ferromagnetic dimers. The difference between experimental and theoretical magnetic structure is perhaps related to the planar magnetic anisotropy in both $BaNd_2ZnO_5$ and $BaNd_2ZnS_5$. Although a weak metamagnetic transition feature was observed in the $BaNd_2ZnO_5$ powder sample [49], no neutron data were available to investigate the corresponding magnetic structure in the metamagnetic state. Since magnetic field is known to introduce different magnetic structures in SS compounds [5,50], more detailed neutron scattering at various applied fields will be necessary to gain physical insights into this family of compounds in the future.

**Conclusion**

We have reported the single crystal growth and anisotropic magnetic characterization of the SS lattice compound. Millimeter-sized single crystals can be obtained through a high temperature solution growth technique. Single crystal structural refinement is consistent with previous powder sample result. The refined anisotropic displacement parameters suggest a small in-plane anisotropy of the Nd ions at room temperature. At high temperatures, $BaNd_2ZnS_5$ is paramagnetic with a large magnetic anisotropy. Local magnetic moments are primarily lying in-plane. A long-range antiferromagnetic ordering is observed at 2.9 K, below which metamagnetic transitions were observed for in-plane [100] and [110] directions. Specific heat measurements show a doublet ground state and a Schottky-like anomaly due to higher lying CEF levels.

**Acknowledgments:** Work done at the University of Arizona was supported by the University of Arizona start-up funds. The research at Oak Ridge National Laboratory (ORNL) was supported by the U.S. Department of Energy (DOE), Office of Science, Office of Basic Energy Sciences, Early Career Research Program Award KC0402020, under Contract DE-AC05-00OR22725.

**Table 1.** Room temperature single crystal refinement information for BaNd$_2$ZnS$_5$.

| Refined Formula | BaNd$_2$ZnS$_5$ |
|---|---|
| FW (g/mol) | 651.49 |
| Space group; Z | $I4/mcm$; 4 |
| a(Å) | 7.83402(16) |
| c(Å) | 13.6071(4) |
| V (Å$^3$) | 835.09(4) |
| Extinction coefficient | 0.0141(4) |
| θ range (°) | 2.994-33.492 |
| No. reflections; R$_{int}$ | 12736; 0.0637 |
| No. independent reflections | 455 |
| No. parameters | 18 |
| R$_1$; ωR$_2$(I>2δ(I)) | 0.0174; 0.0401 |
| Goodness of fit | 1.181 |
| Diffraction peak and hole (e$^-$/Å$^3$) | 1.170, -1.370 |

**Table 2.** Atomic coordinates for BaNd$_2$ZnS$_5$ including the equivalent isotropic displacement parameters.

| Atom | Wyck. | x | y | z | U$_{eq}$ |
|---|---|---|---|---|---|
| Ba1 | 4a | 0 | 0 | ¼ | 0.01247(12) |
| Nd1 | 8h | 0.16230(2) | 0.66230(2) | 0 | 0.00902(11) |
| Zn1 | 4b | 0 | ½ | ¼ | 0.01056(16) |
| S1 | 4c | 0 | 0 | 0 | 0.0112(3) |
| S2 | 16l | 0.65013(7) | 0.15013(7) | 0.13405(7) | 0.01192(17) |

**Table 3.** Anisotropic displacement parameters for each atomic site in BaNd$_2$ZnS$_5$.

| Atom | U11 | U22 | U33 | U23 | U13 | U12 |
|------|-----|-----|-----|-----|-----|-----|
| Ba1 | 0.01218(14) | 0.01218(14) | 0.01305(19) | 0 | 0 | 0 |
| Nd1 | 0.00873(12) | 0.00873(12) | 0.00959(15) | 0 | 0 | -0.00124(6) |
| Zn1 | 0.0120(2) | 0.0120(2) | 0.0077(3) | 0 | 0 | 0 |
| S1 | 0.0093(4) | 0.0093(4) | 0.0149(7) | 0 | 0 | 0 |
| S2 | 0.0125(2) | 0.0125(2) | 0.0107(4) | 0.00212(18) | 0.00212(18) | -0.0015(2) |

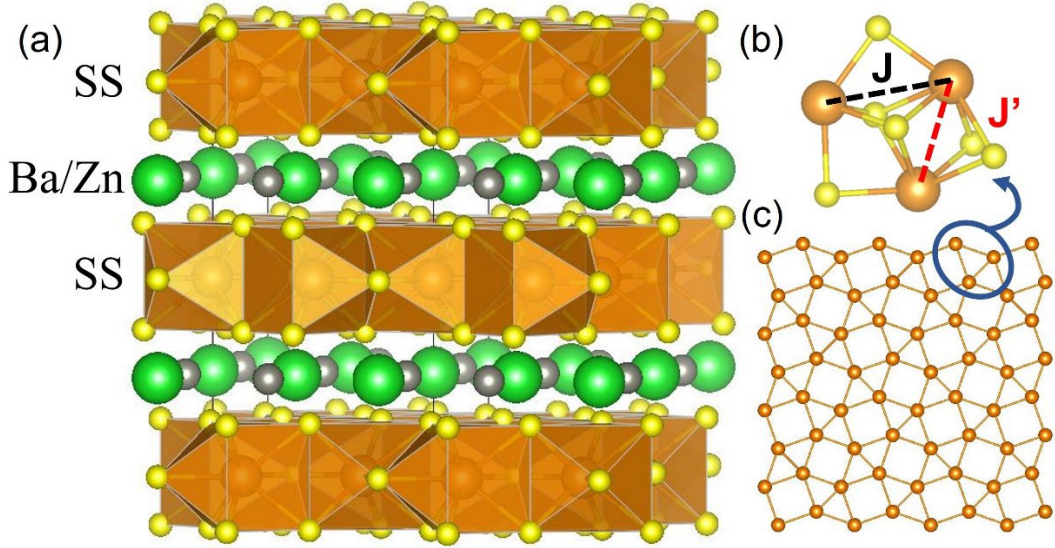

**Figure 1.** (a) Schematic drawing of the crystal structure of BaNd$_2$ZnS$_5$. (b) Nd-S-Nd coordination for the selected triangular motif. (c) Nd sublattice in the *ab*-plane. Barium, zinc, neodymium and sulfur atoms are represented by green, grey, brown and yellow spheres respectively.

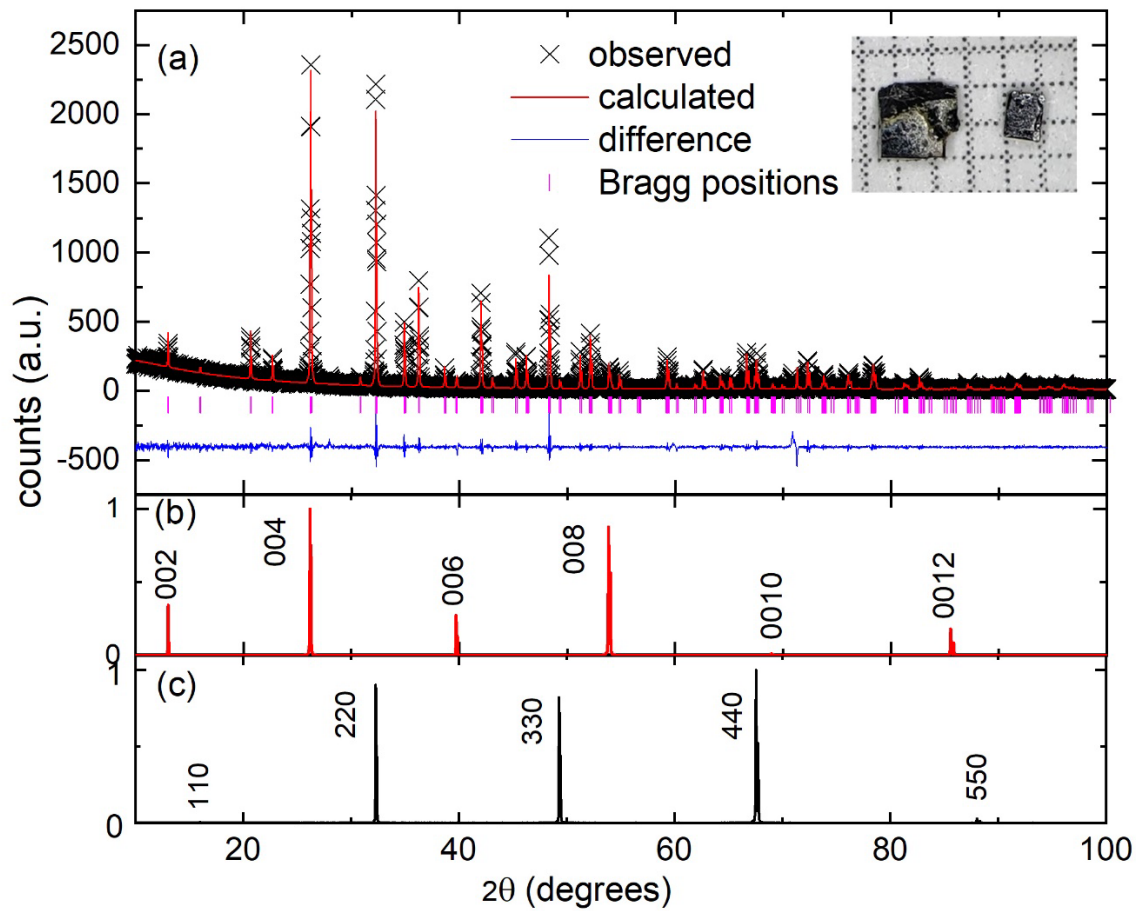

**Figure 2.** (a) Powder x-ray diffraction pattern for BaNdZnS$_5$. Top inset shows as-grown single crystals on a millimeter grid paper. (b) and (c) show diffraction peaks from certain lattice planes.

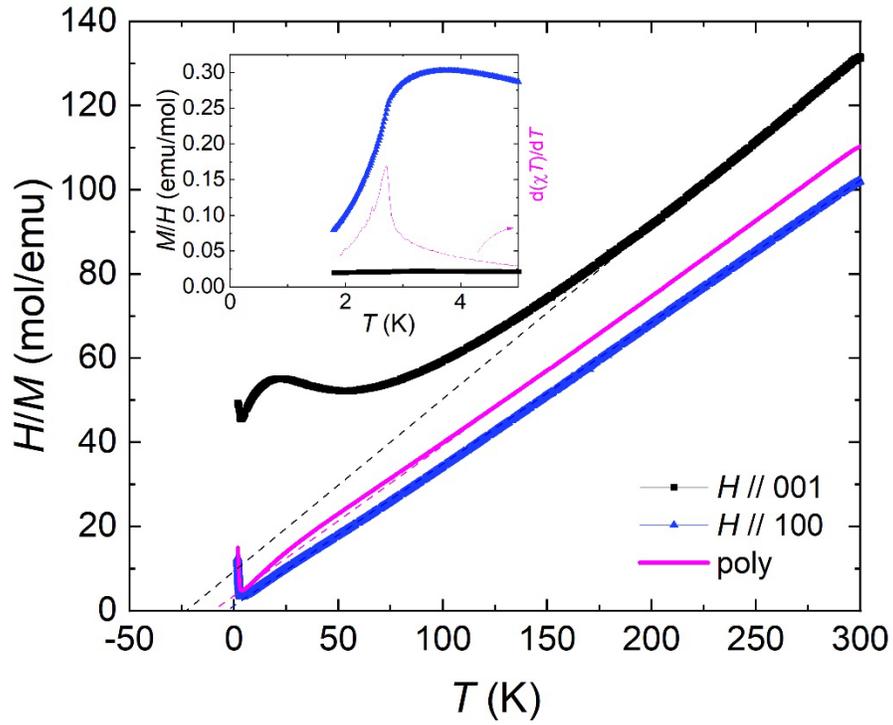

**Figure 3.** Anisotropic temperature-dependent inverse magnetization of BaNd$_2$ZnS$_5$. Dashed lines represent high temperature Curie-Weiss fits. Inset shows zoom-in view of magnetization at low temperatures and d$\chi T$/d$T$ in arbitrary units.

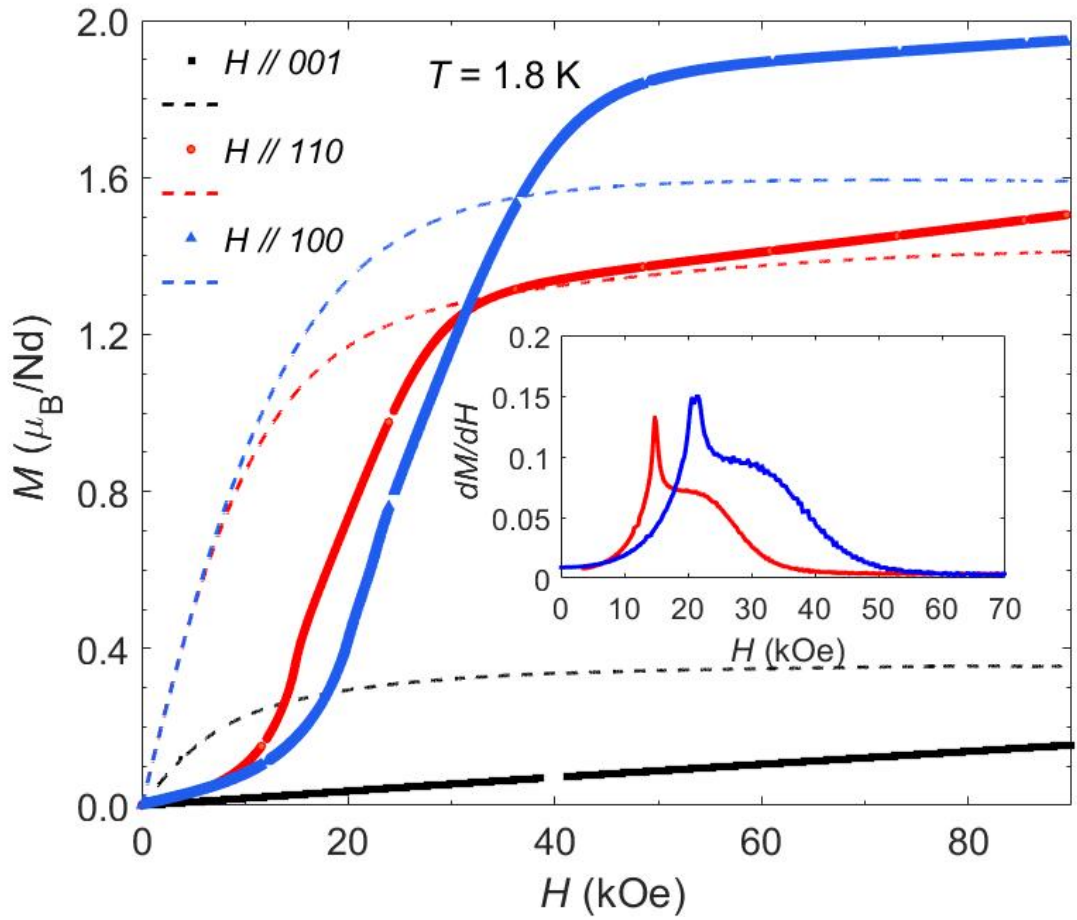

**Figure 4.** Anisotropic magnetic isotherms of BaNd$_2$ZnS$_5$ measured at 1.8 K. Solid lines represent data from pure BaNd$_2$ZnS$_5$. Dashed lines represent data from diluted sample, BaLa$_{1.86}$Nd$_{0.14}$ZnS$_5$, as detailed in the text. Inset shows the d$M$/d$H$ as a function of applied magnetic field up to 70 kOe.

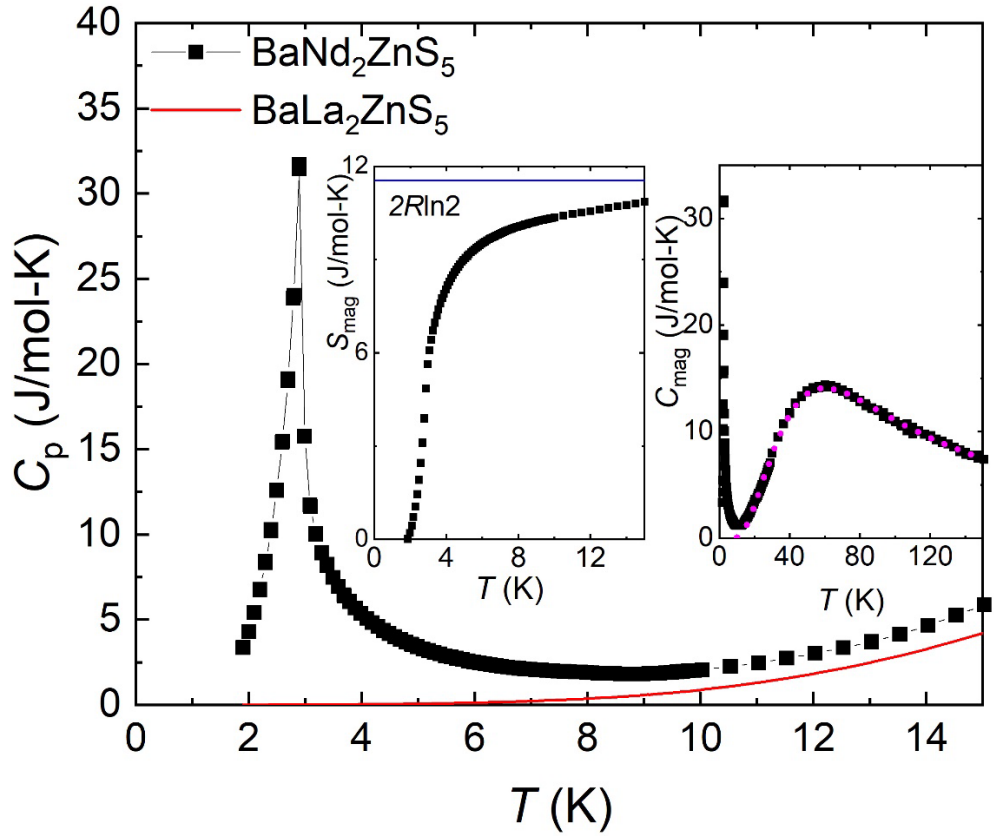

**Figure 5.** Zero-field temperature-dependent specific heat of BaNd$_2$ZnS$_5$. The left inset shows estimated magnetic entropy. The right inset shows high temperature magnetic specific heat with calculated Schottky anomaly contrition (magenta dotted line) as detailed in the text. Red solid line shows the specific heat of BaLa$_2$ZnS$_5$.